\documentclass[aps,prd,twocolumn,superscriptaddress]{revtex4}
\usepackage{epsfig,epsf}
\usepackage{amsmath}
\usepackage{amsthm}
\usepackage{amsfonts}
\usepackage{amssymb}
\usepackage{dsfont}
\usepackage{multirow}
\usepackage{appendix}
\usepackage{slashed}
\usepackage[active]{srcltx}
\usepackage{psfrag}

\begin{document}

\title{Vector resonance $X_1(2900)$ and its structure}
\date{\today}
\author{S.~S.~Agaev}
\affiliation{Institute for Physical Problems, Baku State University, Az--1148 Baku,
Azerbaijan}
\author{K.~Azizi}
\affiliation{Department of Physics, University of Tehran, North Karegar Avenue, Tehran
14395-547, Iran}
\affiliation{Department of Physics, Do\v{g}u\c{s} University, Acibadem-Kadik\"{o}y, 34722
Istanbul, Turkey}
\affiliation{School of Particles and Accelerators, Institute for Research in Fundamental
Sciences (IPM) P.O. Box 19395-5531, Tehran, Iran}
\author{H.~Sundu}
\affiliation{Department of Physics, Kocaeli University, 41380 Izmit, Turkey}

\begin{abstract}
The new vector resonance $X_1(2900)$ observed recently by LHCb in the $%
D^{-}K^{+}$ invariant mass distribution in the decay $B^{+} \to
D^{+}D^{-}K^{+}$ is studied to uncover internal structure of this state, and
calculate its physical parameters. In the present paper, the resonance $%
X_1(2900)$ is modeled as an exotic vector  state, $ J^P=1^- $, built of
the light diquark $u^{T}C\gamma_5d$ and heavy antidiquark $\overline{c}%
\gamma_{\mu}\gamma_5C\overline{s}^{T}$. The mass and current coupling of $%
X_1(2900)$ are computed using the QCD two-point sum rule approach by taking
into account various vacuum condensates up to dimension $10$. The width of
the resonance $X_1(2900)$ is saturated by two decay channels $X_1 \to
D^{-}K^{+}$ and $X_1 \to \overline{D}^{0}K^{0}$. The strong couplings $g_1$
and $g_2$ corresponding to the vertices $X_1D^{-}K^{+}$ and $X_1\overline{D}%
^{0}K^{0}$ are evaluated in the context of the QCD light-cone sum rule
method and technical tools of the soft-meson approximation. Results for the
mass of the resonance $X_1(2900)$ $m=(2890~\pm 122)~\mathrm{MeV}$, and for
its full width $\Gamma _{\mathrm{full}}=(93\pm 13)~\mathrm{MeV}$ are smaller
than their experimental values reported by the LHCb collaboration.
Nevertheless, by taking into account theoretical and experimental errors of
investigations, interpretation of the state $X_1(2900)$ as the vector
tetraquark does not contradict to the LHCb data. We also point out that
analysis of the invariant mass distribution $D^{+}K^{+}$ in the same decay $%
B^{+} \to D^{+}D^{-}K^{+}$ may reveal doubly charged four-quark structures $%
[uc][\overline{s}\overline{d}]$.
\end{abstract}

\maketitle

%%%%%%%%%%%%%%%%%%%%%%%%%%%%%%%%%%%%%%%%%%%%%%%%%%%%%%%%%%%%%%%%%%%%%%%%

\section{Introduction}

\label{sec:Int} %%%%%%%%%%%%%%%%%%%%%%%%%%%%%%%%%%%%%%%%%%%%%%%%%%%%%%%

Recently the LHCb collaboration reported on two new resonant structures $%
X_{0}(2900)$ and $X_{1}(2900)$ (hereafter $X_{0}$ and $X_{1}$) revealed in
the $D^{-}K^{+}$ mass distribution in the process $B^{+}\rightarrow
D^{+}D^{-}K^{+}$ \cite{LHCb:2020A,LHCb:2020}. The collaboration measured the
masses and widths of these structures, as well as determined their
spin-parities. It turned out, that $X_{0}$ and $X_{1}$ are scalar and vector
resonances, respectively, and have close masses. These resonance-like peaks
are first evidence for exotic mesons composed of four quarks of different
flavors provided they can be considered as real resonances. In fact, from
decay channels of $X_{0}$ and $X_{1}$, it is clear that they contain four
valence quarks $ud\overline{s}\overline{c}$, which put these structures to
exclusive place in the $XYZ$ family of exotic mesons.

This discovery triggered interesting theoretical investigations in the
context of various models aimed to account for internal organization of the
resonances $X_{0}$ and $X_{1}$ and calculate their parameters \cite%
{Karliner:2020vsi,Wang:2020xyc,Chen:2020aos,Liu:2020nil,Molina:2020hde,Hu:2020mxp,He:2020jna,Liu:2020orv, Lu:2020qmp,Zhang:2020oze,Huang:2020ptc,Xue:2020vtq, Yang:2021izl,Wu:2020job,Abreu:2020ony,Wang:2020prk,Xiao:2020ltm,Dong:2020rgs,Burns:2020xne,Bondar:2020eoa, Chen:2020eyu,Albuquerque:2020ugi,Chen:2021erj}%
. In these articles various assumptions were made about quark-gluon
structure of the states $X_{0}$ and $X_{1}$, considered their production
mechanisms and decay channels. Investigations were performed in the context
of different models by applying numerous methods and calculational schemes.
Thus, $X_{0}$ was treated as a scalar tetraquark $X_{0}=[sc][\overline{u}%
\overline{d}]$ in Refs.\ \cite{Karliner:2020vsi,Wang:2020xyc}, whereas in
the papers \cite{Chen:2020aos,Liu:2020nil,Molina:2020hde,Hu:2020mxp} the $%
X_{0}$ was interpreted as a scalar molecule $D^{\ast -}K^{\ast +}$. Similar
situation emerges in articles devoted to analysis of the vector resonance $%
X_{1}$. For example, the diquark-antidiquark $X_{1}=[ud][\overline{s}%
\overline{c}]$ and molecule pictures for $X_{1}$ were proposed in Refs. \cite%
{Chen:2020aos,He:2020jna}, respectively. The structures $X_{0}$ and $X_{1}$,
and in general, tetraquarks $\overline{c}\overline{s}qq$ and $sQ\overline{q}%
\overline{q}$ were investigated in the quark models as well \cite%
{Wang:2020prk,Yang:2021izl}.

Recently,  we considered the resonance $X_{0}$ as a bound state of
conventional mesons $\overline{D}^{\ast 0}K^{\ast 0}$ \cite{Agaev:2020nrc}.
We calculated its mass and width and found for these parameters $2868\pm 198~%
\mathrm{MeV}$ and $49.6\pm 9.3~\mathrm{MeV}$, respectively. These results
are in a nice agreement with the LHCb data, which allowed us to classify $%
X_{0}$ as a hadronic molecule state.

However, two enhancements $X_{0}$ and $X_{1}$ in the $D^{-}K^{+}$ mass
distribution may have alternative origin. In fact, the authors of $\ $Ref.
\cite{Liu:2020orv} studied the decay $B^{+}\rightarrow D^{+}D^{-}K^{+}$ via
rescattering diagrams $\chi _{c1}D^{\ast -}K^{\ast +}$ and $D_{sJ}\overline{D%
}_{1}^{0}K^{0}$. It was claimed that, two resonance-like peaks found around
thresholds $D^{\ast -}K^{\ast +}$ and $\overline{D}_{1}^{0}K^{0}$ may
simulate the states $X_{0}$ and $X_{1}$ without a necessity to introduce
genuine four-quark mesons. Appearance of two peaks $X_{0}$ and $X_{1}$ was
explained there by the triangle singularities in the scattering amplitudes
located in the vicinity of the physical boundary.

This rather brief glance at the literature is enough to see that existing
interpretations of $X_{0}$ and $X_{1}$ are controversial and far from being
clear. Although among various approaches diquark-antidiquark and hadronic
molecule models are dominant ones, alternative assumptions deserve detailed
studies as well.

It is worth emphasizing that exotic mesons composed of four quarks of
different flavors already attracted interests of scientists, and valuable
information was collected on their properties. Investigations of such
structures were inspired by observation of the state $X(5568)$, though it
was not confirmed later by other experiments. Nevertheless, performed
analyses led to considerable theoretical progress in understanding of
relevant problems. Indeed, the fully open flavor scalar tetraquark $%
X_{c}=[su][\overline{c}\overline{d}]$ was considered in Refs.\ \cite%
{Agaev:2016lkl} and \cite{Chen:2016mqt}. Doubly charged tetraquarks $Z_{%
\overline{c}s}=[sd][\overline{u}\overline{c}]$ with spin-parities $J^{%
\mathrm{P}}=0^{+},0^{-}$ and $1^{+}$ were explored in our paper \cite%
{Agaev:2017oay}.

The tetraquarks $Z_{\overline{c}s}$ are built of four different quarks as
the resonances $X_{0}$ and $X_{1}$, but bear two units of electric charge
and are objects of special interest. In Ref.\ \cite{Agaev:2017oay} the
masses and full widths of these states were computed in the framework of the
QCD sum rule method. It is instructive to compare parameters of the
axial-vector tetraquark $Z_{\mathrm{AV}}$ with ones of $X_{1}$ although they
are states of different structures and parities. In accordance with our
result, the mass $m_{\mathrm{AV}}=2826_{-157}^{+134}\ \mathrm{MeV}$ of $Z_{%
\mathrm{AV}}$ is comparable with the mass of the state $X_{1}$. But $Z_{%
\mathrm{AV}}$ has the full width $\Gamma _{\mathrm{AV}}=(47.3\pm 11.1)\
\mathrm{MeV}$ and is narrower than the structure $X_{1}$. This is in
contrast to the case of scalar and pseudoscalar tetraquarks $X_{c}$, $Z_{%
\mathrm{S}}$ and $Z_{\mathrm{PS}}$, ground-state masses of which are lower
than the mass of $X_{0}$. Such comparison allows us to conclude that the
vector tetraquark $[ud][\overline{c}\overline{s}]$ has certain chances to
explain observed experimental features of $X_{1}$.

In the present article, we are going to explore the vector resonance $X_{1}$
by assuming that it is a genuine exotic diquark-antidiquark state $%
X_{1}=[ud][\overline{c}\overline{s}]$. Parameters of $X_{1}$ reported by the
LHCb collaboration are:
\begin{eqnarray}
m_{1} &=&(2904\pm 5\pm 1)~\mathrm{MeV},\   \notag \\
\Gamma _{1} &=&(110\pm 11\pm 4)~\mathrm{MeV}.  \label{eq:Data2}
\end{eqnarray}%
We calculate the mass and current coupling of the tetraquark $X_{1}$ by
means of the QCD two-point sum rules. Results obtained for these parameters
are used to evaluate the partial widths of $P$-wave decays $X_{1}\rightarrow
D^{-}K^{+}$ and $X_{1}\rightarrow \overline{D}^{0}K^{0}$ in order to
estimate full width of the resonance $X_{1}$. Our predictions will be
compared with the LHCb data to check validity of suggestions made about a
diquark nature of $X_{1}$.

This paper is organized in the following way: In Section \ref{sec:Masses},
we calculate the mass and coupling of the vector tetraquark $[ud][\overline{c%
}\overline{s}]$. In Section \ref{sec:Decays}, we calculate the strong
couplings $g_{1}$ and $g_{2}$ corresponding to vertices $X_{1}D^{-}K^{+}$
and $X_{1}\overline{D}^{0}K^{0}$. To this end, we use the QCD light-cone sum
rule (LCSR) method and the soft-meson approximation. In this section we find
the partial widths of the processes $X_{1}\rightarrow D^{-}K^{+}$ and $%
X_{1}\rightarrow \overline{D}^{0}K^{0}$. \ Here, the full width of $X_{1}$
is evaluated as well. In Section \ref{sec:Disc}, we discuss obtained
results, propose to study $D^{+}K^{+}$ invariant mass distribution in decay $%
B^{+}\rightarrow D^{+}D^{-}K^{+}$ to observe hypothetical yet doubly charged
scalar and vector tetraquarks $[uc][\overline{s}\overline{d}]$, and conclude
with brief notes.

%%%%%%%%%%%%%%%%%%%%%%%%%%%%%%%%%%%%%%%%%%%%%%%%%%%%%%%%%%%%%%%%%%%%%%%%%%%%

\section{The mass and coupling of $X_{1}$}

\label{sec:Masses}
%%%%%%%%%%%%%%%%%%%%%%%%%%%%%%%%%%%%%%%%%%%%%%%%%%%%%%%%%%%
The mass $m$ and current coupling $f$ of the vector tetraquark $[ud][%
\overline{c}\overline{s}]$ are among key ingredients to check the assumption
about diquark-antidiquark nature of the resonance $X_{1}$. First, the mass
of $X_{1}$ was measured experimentally, therefore prediction obtained for $m
$ should be confronted directly with $m_{1}$. Additionally, the
spectroscopic parameters $m$ and $f$ \ are necessary to find partial widths
of the strong decays $X_{1}\rightarrow D^{-}K^{+}$ \ and $X_{1}\rightarrow
\overline{D}^{0}K^{0}$, and evaluate full width of $X_{1}$.

We compute $m$ and $f$ in the context of the QCD two-point sum rule method
\cite{Shifman:1978bx,Shifman:1978by}. It is one of the effective
nonperturbative approaches to determine parameters of the conventional
hadrons and explore their different decay channels. But this method can also
be applied to study properties of the exotic hadrons. Indeed, the masses and
couplings (or residues) of various tetraquarks, their different decay
channels were investigated within the QCD sum rule method (see, for example,
the review articles \cite%
{Chen:2016qju,Chen:2016spr,Albuquerque:2018jkn,Agaev:2020zad}),

Starting point in our analysis to derive sum rules for $m$ and $f$ is the
two-point correlation function
\begin{equation}
\Pi _{\mu \nu }(p)=i\int d^{4}xe^{ipx}\langle 0|\mathcal{T}\{J_{\mu
}(x)J_{\nu }^{\dag }(0)\}|0\rangle.  \label{eq:CF1}
\end{equation}%
Here, $\mathcal{T}$ stands for the time-ordered product of two currents, and
$J_{\mu }(x)$ is the interpolating current for $X_{1}$. The current for the
vector diquark-antidiquark state $X_{1}$ can be written down in the
following form
\begin{equation}
J_{\mu }(x)=\varepsilon \widetilde{\varepsilon }[u_{b}^{T}(x)C\gamma
_{5}d_{c}(x)][\overline{c}_{m}(x)\gamma _{\mu }\gamma _{5}C\overline{s}%
_{n}^{T}(x)],  \label{eq:CR1}
\end{equation}%
where $\varepsilon \widetilde{\varepsilon }=\varepsilon ^{abc}\varepsilon
^{amn}$, and $a$, $b$, $c$, $m$ and $n$ are color indices. In Eq.\ (\ref%
{eq:CR1}) $u(x)$, $d(x)$, $c(x)$ and $s(x)$ are the quark fields, whereas $C$
denotes the charge-conjugation operator. The  isoscalar current  $J_{\mu }(x)$ for the
tetraquark $X_{1}$ with quantum numbers $J^{\mathrm{P}}=1^{-}$ is built of
the light scalar diquark $uC\gamma _{5}d$ and heavy vector antidiquark $%
\overline{c}\gamma _{\mu }\gamma _{5}C\overline{s}$, which belongs to
antitriplet and triplet representations of the color group, respectively. As
a result, the interpolating current $J_{\mu }(x)$ belongs to $[\overline{%
\mathbf{3}}_{c}]\otimes \lbrack \mathbf{3}_{c}]$ representation of $%
SU_{c}(3) $, and is a colorless construction.

To derive desired sum rules for parameters of $X_{1}$, we have to represent
the correlation function $\Pi _{\mu \nu }(p)$ in terms of these parameters,
and get the phenomenological side of the sum rules $\Pi _{\mu \nu }^{\mathrm{%
Phys}}(p)$. In terms of the tetraquark's parameters the correlation function
has the following form
\begin{equation}
\Pi _{\mu \nu }^{\mathrm{Phys}}(p)=\frac{\langle 0|J_{\mu }|X_{1}\rangle
\langle X_{1}|J_{\nu }^{\dagger }|0\rangle }{m^{2}-p^{2}}+\cdots.
\label{eq:CF2}
\end{equation}%
Expression (\ref{eq:CF2}) is obtained by saturating the correlation function
$\Pi _{\mu \nu }(p)$ with a complete set of $J^{\mathrm{P}}=1^{-}$ states
and performing integration over $x$ in Eq.\ (\ref{eq:CF1}): Contributions of
higher resonances and continuum states in $X_{1}$ channel are shown by dots.

The correlator $\Pi _{\mu \nu }^{\mathrm{Phys}}(p)$ can be detailed by
introducing the matrix element
\begin{equation}
\langle 0|J_{\mu }|X_{1}\rangle =fm\epsilon _{\mu },  \label{eq:MElem1}
\end{equation}%
where $\epsilon _{\mu }$ is the polarization vector of the state $X_{1}$.
Then $\Pi _{\mu \nu }^{\mathrm{Phys}}(p)$ takes the following form
\begin{equation}
\Pi _{\mu \nu }^{\mathrm{Phys}}(p)=\frac{m^{2}f^{2}}{m^{2}-p^{2}}\left(
-g_{\mu \nu }+\frac{p_{\mu }p_{\nu }}{m^{2}}\right) +\cdots ,
\label{eq:CorF1}
\end{equation}%
and contains  in parentheses the Lorentz structure of the vector state. A
part of this structure proportional to $g_{\mu \nu }$ receives contribution
only from vector states, therefore in our analysis we use this term and
corresponding invariant amplitude $\Pi ^{\mathrm{Phys}}(p^{2})$.

We approximate the phenomenological side of the sum rule $\Pi _{\mu \nu }^{%
\mathrm{Phys}}(p)$ in Eq.\ (\ref{eq:CF2}) using a simple-pole term. For
multiquark hadrons such treatment may give rise to some doubts, because $\Pi
_{\mu \nu }^{\mathrm{Phys}}(p)$ contains also contributions of two-hadron
reducible terms. Indeed, the relevant interpolating current interacts not
only to a multiquark hadron, but couples also with two conventional hadrons
lying below the mass of the multiquark system \cite{Kondo:2004cr,Lee:2004xk}%
. Such two-hadron states generate the finite width $\Gamma (p)$ of the
multiquark hadron, and modify the quark propagator
\begin{equation}
\frac{1}{m^{2}-p^{2}}\rightarrow \frac{1}{m^{2}-p^{2}-i\sqrt{p^{2}}\Gamma (p)%
}.  \label{eq:Mod}
\end{equation}%
In the case of the tetraquark, these effects rescale its coupling $f$
leaving fixed the mass $m$. Detailed analyses demonstrated that two-hadron
contributions as a whole, and two-meson ones in particular are small, and
can be neglected \cite{Lee:2004xk,Wang:2015nwa,Agaev:2018vag,Sundu:2018nxt}.
Therefore, in Eq.\ (\ref{eq:CF2}) we use the zero-width single-pole
approximation.

The QCD side of the sum rules $\Pi _{\mu \nu }^{\mathrm{OPE}}(p)$ should be
computed in the operator product expansion ($\mathrm{OPE}$) with some
accuracy. To get $\Pi _{\mu \nu }^{\mathrm{OPE}}(p)$, we insert into Eq.\ (%
\ref{eq:CF1}) the interpolating current $J(x)$, and contract relevant heavy
and light quark fields. After these operations, for $\Pi ^{\mathrm{OPE}}(p)$
we find
\begin{eqnarray}
&&\Pi _{\mu \nu }^{\mathrm{OPE}}(p)=i\int d^{4}xe^{ipx}\varepsilon
\widetilde{\varepsilon }\varepsilon ^{\prime }\widetilde{\varepsilon }%
^{\prime }\mathrm{Tr}\left[ \gamma _{5}\widetilde{S}_{u}^{bb^{\prime
}}(x)\gamma _{5}S_{d}^{cc^{\prime }}(x)\right]  \notag \\
&&\times \mathrm{Tr}\left[ \gamma _{\mu }\gamma _{5}\widetilde{S}%
_{s}^{n^{\prime }n}(-x)\gamma _{5}\gamma _{\nu }S_{c}^{m^{\prime }m}(-x)%
\right],  \label{eq:QCD1}
\end{eqnarray}%
where $ \widetilde{S}_{c(q)}(x) =CS^T_{c(q)}C$.
Here, $S_{c}(x)$ and $S_{u(s,d)}(x)$ are the heavy $c$- and light $u(s,d)$%
-quark propagators, respectively. Their explicit expressions are presented,
for example, in Ref.\ \cite{Agaev:2020zad}. We denote by $\Pi ^{\mathrm{OPE}%
}(p^{2})$ the invariant amplitude corresponding to the structure $g_{\mu \nu
}$ in Eq.\ (\ref{eq:QCD1}), and use it in our following investigations.

The sum rules for the parameters $m$ and $f$ can be found by equating $\Pi ^{%
\mathrm{Phys}}(p^{2})$ and $\Pi ^{\mathrm{OPE}}(p^{2})$ and carrying out
usual operations necessary in QCD sum rule computations. These operations
include the Borel transformation of the invariant amplitudes and subtraction
higher resonances and continuum terms from the phenomenological side using
the assumption on the quark-hadron duality. After these manipulations, the
sum rules acquire dependence on the Borel $M^{2}$ and continuum threshold $%
s_{0}$ parameters.

The sum rules for $m$ and $f$ read
\begin{equation}
m^{2}=\frac{\Pi ^{\prime }(M^{2},s_{0})}{\Pi (M^{2},s_{0})},  \label{eq:Mass}
\end{equation}%
and
\begin{equation}
f^{2}=\frac{e^{m^{2}/M^{2}}}{m^{2}}\Pi (M^{2},s_{0}).  \label{eq:Coupl}
\end{equation}%
Here, $\Pi (M^{2},s_{0})$ is the Borel transformed and subtracted invariant
amplitude $\Pi ^{\mathrm{OPE}}(p^{2})$, and $\Pi ^{\prime
}(M^{2},s_{0})=d/d(-1/M^{2})\Pi (M^{2},s_{0})$.

The function $\Pi (M^{2},s_{0})$ has the following form%
\begin{equation}
\Pi (M^{2},s_{0})=\int_{\mathcal{M}^{2}}^{s_{0}}ds\rho ^{\mathrm{OPE}%
}(s)e^{-s/M^{2}}+\Pi (M^{2}).  \label{eq:InvAmp}
\end{equation}%
In this paper we neglect the mass of the $u$ and $d$
quarks, therefore in Eq.\ (\ref{eq:InvAmp}) $\mathcal{M}=m_{c}+m_{s}$. The spectral density $\rho ^{\mathrm{OPE}}(s)$ is found as an imaginary part
of the correlation function and encompasses essential piece of $\Pi ^{%
\mathrm{OPE}}_{\mu\nu}(p)$.  The
Borel transformations of remaining terms in $\Pi ^{\mathrm{OPE}}_{\mu\nu}(p)$ are
included into $\Pi (M^{2})$: the latter was calculated directly from the
expression of $\Pi ^{\mathrm{OPE}}_{\mu\nu}(p)$.

Computations are performed by including into analysis vacuum condensates up
to dimension $10$. We use the basic quark, gluon and mixed condensates, as
well as higher ones obtained as their products: We assume that the
factorization of higher dimensional contributions does not generate large
ambiguities. We do not provide here analytical expressions of $\rho ^{%
\mathrm{OPE}}(s)$ and $\Pi (M^{2})$, because they are rather lengthy.

The numerical values of the basic condensates were extracted from analysis
of different hadronic processes, and are well known parameters \cite%
{Shifman:1978bx,Shifman:1978by,Ioffe:2005ym}
\begin{eqnarray}
&&\langle \overline{q}q\rangle =-(0.24\pm 0.01)^{3}~\mathrm{GeV}^{3},\
\langle \overline{s}s\rangle =(0.8\pm 0.1)\langle \overline{q}q\rangle ,
\notag \\
&&\langle \overline{q}g_{s}\sigma Gq\rangle =m_{0}^{2}\langle \overline{q}%
q\rangle ,\ \langle \overline{s}g_{s}\sigma Gs\rangle =m_{0}^{2}\langle
\overline{s}s\rangle ,  \notag \\
&&m_{0}^{2}=(0.8\pm 0.2)~\mathrm{GeV}^{2}  \notag \\
&&\langle \frac{\alpha _{s}G^{2}}{\pi }\rangle =(0.012\pm 0.004)~\mathrm{GeV}%
^{4},  \notag \\
&&\langle g_{s}^{3}G^{3}\rangle =(0.57\pm 0.29)~\mathrm{GeV}^{6}.
\label{eq:Parameters}
\end{eqnarray}%
For the gluon condensate $\langle g^{3}G^{3}\rangle $ we employ the estimate
given in Ref.\ \cite{Narison:2015nxh}. The QCD sum rules contain also $c$
and $s$ quark masses for which we use $m_{s}=93_{-5}^{+11}~\mathrm{MeV}$,
and $\ m_{c}=1.27\pm 0.2~\mathrm{GeV}$.

The sum rules for $m$ and $f$ depend also on the auxiliary parameters of
computations, i.e., are functions of $M^{2}$ and $s_{0}$. The working
regions for $M^{2}$ and $s_{0}$ should meet usual requirements imposed on
the pole contribution ($\mathrm{PC}$) and convergence of the operator
product expansion. We explore $\mathrm{PC}$ and convergence of $\mathrm{OPE}$
by means of the quantities
\begin{equation}
\mathrm{PC}=\frac{\Pi (M^{2},s_{0})}{\Pi (M^{2},\infty )},  \label{eq:PC}
\end{equation}%
and
\begin{equation}
R=\frac{\Pi ^{\mathrm{DimN}}(M^{2},s_{0})}{\Pi (M^{2},s_{0})}.
\label{eq:Convergence}
\end{equation}%
In Eq.\ (\ref{eq:Convergence}) $\Pi ^{\mathrm{DimN}}(M^{2},s_{0})$ is a last
term (or a sum of last few terms) in the correlation function. In the
present article, we use the sum of last three terms in $\mathrm{OPE}$, and $%
\mathrm{DimN\equiv Dim(8+9+10)}$.

The $\mathrm{PC}$ is necessary to determine the upper limit for $M^{2}$,
whereas the lower bound for the Borel parameter is fixed from analysis of $R$%
. These two limits for $M^{2}$ determine boundaries of the working window,
where the Borel parameter can be varied. Our calculations prove that the
regions for the parameters $M^{2}$ and $s_{0}$ are
\begin{equation}
M^{2}\in \lbrack 3,3.5]\ \mathrm{GeV}^{2},\ s_{0}\in \lbrack 11.3,13.3]\
\mathrm{GeV}^{2}.  \label{eq:Wind1}
\end{equation}%
These regions obey standard restrictions on $\mathrm{PC}$ and convergence of
$\mathrm{OPE}$. In fact, at $M^{2}=3~\mathrm{GeV}^{2}$ the pole contribution
is $0.7$, whereas at $M^{2}=3.5~\mathrm{GeV}^{2}$ is equal to $0.21$. At the
minimum of $M^{2}=3~\mathrm{GeV}^{2}$, we find $R\approx 0.01$ which
indicates about the convergence of the sum rules. The parameters $m$ and $f$
\ are extracted approximately at a middle of the window (\ref{eq:Wind1}),
i.e., at $M^{2}=3.25~\mathrm{GeV}^{2}$ and $s_{0}=12~\mathrm{GeV}^{2}$,
where $\mathrm{PC}\approx 0.55$ ensuring the ground state nature of $X_{1}$.

Our predictions for $m$ and $f$ are
\begin{eqnarray}
m &=&(2890~\pm 122)~\mathrm{MeV},  \notag \\
f &=&(2.1\pm 0.4)\times 10^{-3}~\mathrm{GeV}^{4}.  \label{eq:Result1}
\end{eqnarray}%
Dependence of the spectroscopic parameters $m$ and $f$ on the choice of $%
M^{2}$ generate an important part of theoretical uncertainties shown in Eq.\
(\ref{eq:Result1}). In the case of $m$ these uncertainties are equal to $\pm
4.2\%$, whereas for the coupling $f$ they amount to $\pm 19.1\%$.
Theoretical uncertainties for the mass are considerably smaller than that
for the coupling, because $m$ is determined by the ratio of correlation
functions, and is exposed to smaller variations. Because the coupling $f$
depends directly on $\Pi (M^{2},s_{0})$ uncertainties are considerably
larger, nevertheless, even in this case they do not exceed limits accepted
in sum rule computations.

\begin{widetext}

\begin{figure}[h!]
\begin{center}
\includegraphics[totalheight=6cm,width=8cm]{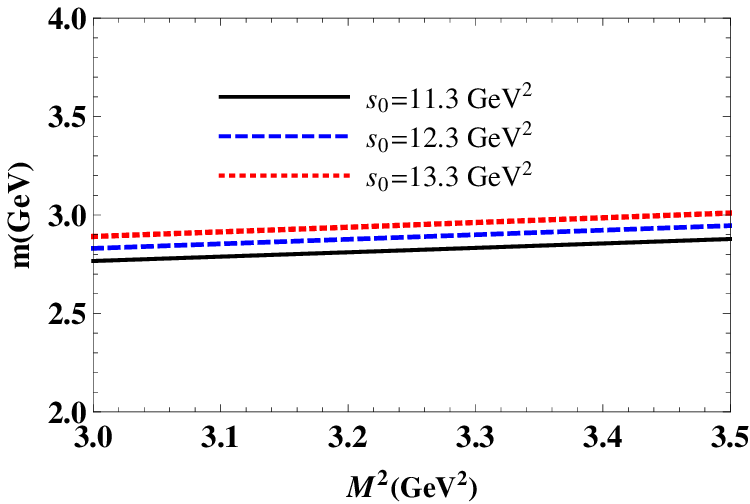}
\includegraphics[totalheight=6cm,width=8cm]{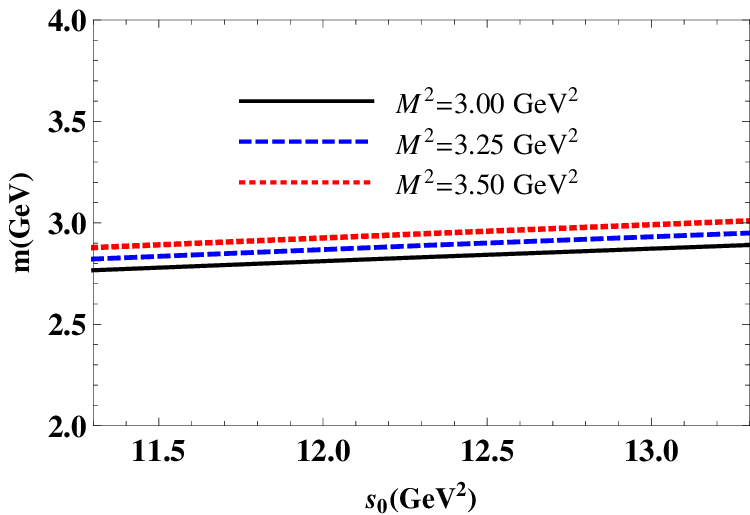}
\end{center}
\caption{The mass $m$ of the tetraquark $X_1$ as a function of the Borel parameter $M^{2}$ (left panel), and as a function of the continuum threshold parameter $s_0$ (right panel).}
\label{fig:Mass}
\end{figure}

\end{widetext}

The continuum threshold parameter $s_{0}$ separates a ground-state
contribution from effects due to higher resonances and continuum states, and
$\sqrt{s_{0}}$ has to be smaller than the mass $m^{\ast }$ of the first
excitation of $X_{1}$. The self-consistent analysis implies that a
difference $m-\sqrt{s_{0}}$ should be around of $m^{\ast }-m$. The mass gap $%
\sqrt{s_{0}}-m\approx (500-600)~\mathrm{MeV}$ fixed in the present work can
be considered as a reasonable estimate $m^{\ast }\approx (m+500)~\mathrm{MeV}
$ for the tetraquark $X_{1}$.

In Fig.\ \ref{fig:Mass} (left panel), we display the sum rule's results for
the mass as a function of $M^{2}$, where one can see a residual dependence
of $m$ on the Borel parameter. A sensitivity of $m$ to the continuum
threshold parameter $s_{0}$ is depicted in right panel of this figure.

Obtained prediction for the mass of the state $X_{1}$ is compatible with the
LHCb data given in Eq.  (\ref{eq:Data2}). But this information is not enough to make a
reliable conclusion about diquark-antidiquark structure of $X_{1}$. For this
purpose, we have to compute the full width $\Gamma _{\mathrm{full}}$ of the
tetraquark $X_{1}$, and compare it with the relevant LHCb data. Only
together the parameters $m$ and $\Gamma _{\mathrm{full}}$ can confirm or not
the assumption about the nature of $X_{1}$.

%%%%%%%%%%%%%%%%%%%%%%%%%%%%%%%%%%%%%%%%%%%%%%%%%%%%%%%%%%%%%%%%%%%%%%%%%%%%

\section{Partial widths of the decays $X_{1}\rightarrow D^{-}K^{+}$ and $X_{1}\rightarrow \overline{D}^{0}K^{0}$}

\label{sec:Decays}
%%%%%%%%%%%%%%%%%%%%%%%%%%%%%%%%%%%%%%%%%%%%%%%%%%%%%%%%%%%

The full width of the tetraquark $X_{1}$ is the sum of partial widths of its
different decay channels. The resonance $X_{1}$ was observed in the $%
D^{-}K^{+}$ mass distribution, therefore we consider the process $%
X_{1}\rightarrow D^{-}K^{+}$ as its main decay channel. There is also
another channel $X_{1}\rightarrow \overline{D}^{0}K^{0}$, which contributes
to full width of $X_{1}$. It is worth noting that these two decays are $P$%
-wave processes. In $S$-wave the tetraquark $X_{1}$ may decay to a pair of
either vector and scalar or axial-vector and pseudoscalar $D$ and $K$
mesons. Therefore, it is not difficult to fix possible $S$-wave channels of $%
X_{1}$. Thus, decays to meson pairs $\overline{D}_{1}(2420)^{0}K^{0}$, $%
\overline{D}_{0}^{\ast }(2400)^{0}K^{*}(892)^{0}$, $\overline{D}^{0}K_{1}(1270)^{0},$ $%
D^{-}K_{1}(1270)^{+}$ and a few other processes would be such channels.
Threshold $2919~\mathrm{MeV}$ for production of a pair $\overline{D}%
_{1}(2420)^{0}K^{0}$ is very close to the mass of the resonance $X_{1}$, and
is lowest one among listed decays. But even this threshold exceeds the mass
of $X_{1}$ making kinematically forbidden the decay $X_{1}\rightarrow
\overline{D}_{1}(2420)^{0}K^{0}$ and other $S$-wave processes.

In this section, we investigate the strong decay $X_{1}\rightarrow
D^{-}K^{+} $ and calculate its partial width. Here, we provide details of
these calculations, but will write down only final predictions for the
second channel $X_{1}\rightarrow \overline{D}^{0}K^{0}$.

The width of the decay $X_{1}\rightarrow D^{-}K^{+}$ is determined by the
strong coupling $g_{1}$ corresponding to the vertex $X_{1}D^{-}K^{+}$. We
compute $g_{1}$ using the QCD light-cone sum rule method \cite%
{Balitsky:1989ry,Belyaev:1994zk} and technical tools known as a soft-meson
approximation \cite{Ioffe:1983ju}. Starting quantity in this method is the
correlation function
\begin{equation}
\Pi _{\mu }(p,q)=i\int d^{4}xe^{ipx}\langle K(q)|\mathcal{T}\{J^{D}(x)J_{\mu
}^{\dag }(0)\}|0\rangle ,  \label{eq:CorrF3}
\end{equation}%
where by $K$ and $D$ we denote the mesons $K^{+}$ and $D^{-}$, respectively.
In the correlation function $\Pi _{\mu }(p,q)$, the interpolating current $%
J_\mu (x)$ is given by Eq.\ (\ref{eq:CR1}). For $J^{D}(x)$ we use
\begin{equation}
J^{D}(x)=\overline{c}_{l}(x)i\gamma _{5}d_{l}(x),  \label{eq:Dcur}
\end{equation}%
where $l$ is the color index.

The function $\Pi _{\mu }(p,q)$ has to be rewritten in terms of the physical
parameters of the initial and final particles involved into the decay. By
taking into account the ground states in the $D$ and $X_{1}$ channels, we
get
\begin{eqnarray}
\Pi _{\mu }^{\mathrm{Phys}}(p,q)&=&\frac{\langle 0|J^{D}|D\left( p\right)
\rangle }{p^{2}-m_{D}^{2}}\langle D\left( p\right) K(q)|X_{1}(p^{\prime
})\rangle  \notag \\
&&\times \frac{\langle X_{1}(p^{\prime })|J_{\mu }^{\dagger }|0\rangle }{%
p^{\prime 2}-m^{2}}+\cdots ,  \label{eq:CorrF4}
\end{eqnarray}%
where $p$, $q$ and $p^{\prime }=p+q$ are the momenta of the particles $D$, $%
K $, and $X_{1}$, respectively. In Eq.\ (\ref{eq:CorrF4}) $m_{D}$ is the
mass of $D^{-}$ meson, and the ellipses refer to contributions of higher
resonances and continuum states in the $D$ and $X_{1}$ channels.

To continue calculations of $\Pi _{\mu }^{\mathrm{Phys}}(p,q)$, we introduce
the matrix elements
\begin{eqnarray}
&&\langle 0|J^{D}|D\rangle =\frac{f_{D}m_{D}^{2}}{m_{c}},\ \langle
X_{1}(p^{\prime })|J_{\mu }^{\dagger }|0\rangle =fm\epsilon _{\mu }^{\ast },
\notag \\
&&\langle D\left( p\right) K(q)|X_{1}(p^{\prime })\rangle =g_1 p\cdot \epsilon .
\label{eq:Mel}
\end{eqnarray}%
In expressions above $f_{D}\,$ is decay constant of the meson $D^{-}$.
Having inserted these matrix elements into expression of the correlation
function, we get for $\Pi _{\mu }^{\mathrm{Phys}}(p,q)$
\begin{eqnarray}
&&\Pi _{\mu }^{\mathrm{Phys}}(p,q)=\frac{g_{1}f_{D}fm_{D}^{2}}{%
2m_{c}m(p^{2}-m_{D}^{2})(p^{\prime 2}-m^{2})}  \notag \\
&&\times \left[ \left( m_{D}^{2}-m_{K}^{2}-m^{2}\right) p_{\mu }+\left(
m^{2}+m_{D}^{2}-m_{K}^{2}\right) q_{\mu }\right] +\cdots,\nonumber\\  \label{eq:CorrF5}
\end{eqnarray}%
with $m_{K}$ being the mass of the $K^{+}$ meson. The function $\Pi _{\mu }^{%
\mathrm{Phys}}(p,q)$ contains two structures proportional to $p_{\mu }$ and $%
q_{\mu }$, respectively. Both of them can be used to derive the sum rule for
the strong coupling $g_{1}$. In what follows, we work with the structure $%
\sim q_{\mu }$ and corresponding invariant amplitude $\Pi ^{\mathrm{Phys}%
}(p^{2})$.

We also have to compute $\Pi _{\mu }(p,q)$ by means of the quark
propagators, and find the QCD side of the sum rule. Contractions of
corresponding quark and antiquark fields in Eq.\ (\ref{eq:CorrF3}) yield
\begin{eqnarray}
\Pi _{\mu }^{\mathrm{OPE}}(p,q)&=&\int d^{4}xe^{ipx}\varepsilon \widetilde{%
\varepsilon }\left[ \gamma _{5}\widetilde{S}_{d}^{lb}(x){}\gamma _{5}\right.
\left.  \widetilde{S}_{c}^{ml}(-x){}\gamma _{\mu }\gamma _{5}\right]
_{\alpha \beta }\notag \\ &&\times\langle K(q)|\overline{u}_{\alpha }^{c}(0)s_{\beta
}^{n}(0)|0\rangle ,  \label{eq:CorrF6}
\end{eqnarray}%
where $\alpha $ and $\beta $ are the spinor indexes.

The function $\Pi ^{\mathrm{OPE}}(p,q)$ contains local matrix elements of
the quark operator $\overline{u}s$ sandwiched between the vacuum and $K$
meson. After some manipulations $\langle K(q)|\overline{u}_{\alpha
}^{c}(0)s_{\beta }^{n}(0)|0\rangle $ can be expressed in terms of the $K$
meson local matrix elements. To this end, $\overline{u}(0)s(0)$ should be
expanded over the full set of Dirac matrices $\Gamma ^{J}$ and projected
onto the color-singlet states
\begin{equation}
\overline{u}_{\alpha }^{c}(0)s_{\beta }^{n}(0)\rightarrow \frac{1}{12}\delta
^{cn}\Gamma _{\beta \alpha }^{J}\left[ \overline{u}(0)\Gamma ^{J}s(0)\right]
,  \label{eq:MatEx}
\end{equation}%
where
\begin{equation}
\Gamma ^{J}=\mathbf{1},\ \gamma _{5},\ \gamma _{\mu },\ i\gamma _{5}\gamma
_{\mu },\ \sigma _{\mu \nu }/\sqrt{2}.  \label{eq:Dirac}
\end{equation}%
Then colorless operators $\overline{u}(0)\Gamma ^{J}s(0)$ give rise to
matrix elements of the $K$ meson.

The expression (\ref{eq:CorrF6}) demonstrates a difference between vertices
of ordinary mesons and ones composed of a tetraquark and two conventional
mesons. Thus, the vertices of ordinary mesons contain non-local matrix
elements $\overline{q}_{1}(x)\Gamma ^{J}q_{2}(0)$ which are connected with
distribution amplitudes (DAs) of a final-state meson. In the case under
consideration, instead of DAs of the $K$ meson, $\Pi ^{\mathrm{QCD}}(p,q)$
depends on its local matrix elements. This difference is generated by the
structure of the interpolating current $J_{\mu }(x)$, which is built of four
quark fields at the same space-time point. Therefore, after contracting
relevant quark fields in $\Pi ^{\mathrm{QCD}}(p,q)$ remaining two quarks of $%
X_{1}$ constitute local matrix elements of the $K$ meson. As a result,
standard integrals over DAs reduce to overall normalization factors. In the
context of the LCSR method this is possible in the limit $q\rightarrow 0$,
when the light-cone expansion is replaced by the short-distant one \cite%
{Belyaev:1994zk}. In this approximation $p=p^{\prime }$ and invariant
amplitudes $\Pi ^{\mathrm{Phys}}(p^{2})$ and $\Pi ^{\mathrm{OPE}}(p^{2})$
depend only on one variable $p^{2}$. The limit $q\rightarrow 0$ is known as
the soft-meson approximation to full light-cone expressions. For our
purposes important is the observation made in Ref.\ \cite{Belyaev:1994zk}:
the soft-meson approximation and full LCSR treatment of the conventional
mesons' vertices lead to results, which are very close to each other.

It is clear, that in this approximation the QCD side of the LCSR becomes
simpler than in its full version. But soft-meson approach gives rise to
complications in the phenomenological side of the sum rule. Thus, in the
soft limit we get for the amplitude $\Pi ^{\mathrm{Phys}}(p^{2})$
\begin{eqnarray}
\Pi ^{\mathrm{Phys}}(p^{2}) &=&g_{1}\frac{f_{D}fm_{D}^{2}}{2mm_{c}(p^{2}-%
\widetilde{m}^{2})^{2}}  \notag \\
&&\times \left( 2\widetilde{m}^{2}-m_{K}^{2}\right) +\cdots ,
\label{eq:CF2a}
\end{eqnarray}%
where $\widetilde{m}^{2}=(m^{2}+m_{D}^{2})/2$. This amplitude contains the
double pole at $p^{2}=\widetilde{m}^{2}$, and its Borel transformation is
given by the formula
\begin{eqnarray}
&&\Pi ^{\mathrm{Phys}}(M^{2})=g_{1}\frac{f_{D}fm_{D}^{2}}{2mm_{c}}\left( 2%
\widetilde{m}^{2}-m_{K}^{2}\right)  \notag \\
&&\times \frac{e^{-\widetilde{m}^{2}/M^{2}}}{M^{2}}+\cdots .  \label{eq:CF3a}
\end{eqnarray}

Apart from ground-state contribution, in the soft limit the amplitude $\Pi ^{%
\mathrm{Phys}}(M^{2})$ contains additional unsuppressed terms. In other
words, double Borel transformation could not suppress all required terms.
These contaminating contributions can be removed from the phenomenological
side of the sum rule by applying to $\Pi ^{\mathrm{Phys}}(M^{2})$ the
operator \cite{Belyaev:1994zk,Ioffe:1983ju}
\begin{equation}
\mathcal{P}(M^{2},m^{2})=\left( 1-M^{2}\frac{d}{dM^{2}}\right)
M^{2}e^{m^{2}/M^{2}}.  \label{eq:Oper}
\end{equation}%
Contribution of terms remained in $\Pi ^{\mathrm{Phys}}(M^{2})$ after this
operation can be subtracted by a usual manner. Naturally, one should act by
operator $\mathcal{P}(M^{2},m^{2})$ also to QCD side of the sum rule. Then,
the sum rule for the strong coupling $g_{1}$ is given by expression%
\begin{equation}
g_{1}=\frac{2mm_{c}(2\widetilde{m}^{2}-m_{K}^{2})}{f_{D}fm_{D}^{2}}\mathcal{P%
}(M^{2},\widetilde{m}^{2})\Pi ^{\mathrm{OPE}}(M^{2},s_{0}),
\label{eq:SRcoupl}
\end{equation}%
where $\Pi ^{\mathrm{OPE}}(M^{2},s_{0})$ is Borel transformed and subtracted
invariant amplitude corresponding to the structure $q_{\mu }$ in $\Pi _{\mu
}^{\mathrm{OPE}}(p,q)$.

Procedures to calculate the correlation function $\Pi _{\mu }^{\mathrm{OPE}%
}(p,q)$ in the soft approximation were presented in Refs.\ \cite%
{Agaev:2016ijz,Agaev:2016dev}, therefore we provide only important points of
these computations. First of all, after substituting the expansion (\ref%
{eq:MatEx}) into Eq.\ (\ref{eq:CorrF3}), one carries out summations over
color indices and determines local matrix elements of the $K$ meson that
contribute to $\Pi _{\mu }^{\mathrm{OPE}}(p,q)$ in the soft-meson
approximation. There is limited number of matrix elements that may
contribute to the correlation function. They are two-particle matrix
elements of twist-2 and twist-3
\begin{eqnarray}
&&\langle 0|\overline{u}\gamma _{\mu }\gamma _{5}s|K(q)\rangle =if_{K}q_{\mu
},  \notag \\
&&\langle 0|\overline{u}i\gamma _{5}s|K\rangle =\frac{f_{K}m_{K}^{2}}{m_{s}}.
\label{eq:MatElK1}
\end{eqnarray}%
as well as three-particle local matrix elements of $K$ meson, for an example,%
\begin{equation}
\langle 0|\overline{u}\gamma ^{\nu }\gamma _{5}igG_{\mu \nu }s|K(q)\rangle
=iq_{\mu }f_{K}m_{K}^{2}\kappa _{4K},  \label{eq:MatElK2}
\end{equation}%
where $f_{K}$ and $\kappa _{4K}$ are the decay constant and the twist-4
matrix element of the $K$ meson. It turns out that in the soft limit
contributions to $\Pi _{\mu }^{\mathrm{OPE}}(p,q)$ comes from the matrix
elements (\ref{eq:MatElK1}). First of them contributes to the structure $%
\sim q_{\mu }$, whereas $\langle 0|\overline{u}i\gamma _{5}s|K\rangle $
forms the second component of $\Pi _{\mu }^{\mathrm{OPE}}(p,q)$ proportional
to $p_{\mu }$.

It has been emphasized above that, we consider the first component of the
correlation function $\Pi _{\mu }^{\mathrm{OPE}}(p,q)$. For the structure $%
\sim q_{\mu }$ the amplitude $\Pi ^{\mathrm{OPE}}(M^{2},s_{0})$ is given by
the expression
\begin{eqnarray}
\Pi ^{\mathrm{OPE}}(M^{2},s_{0}) &=&-\frac{f_{K}}{48\pi ^{2}}\int_{\mathcal{M%
}^{2}}^{s_{0}}\frac{ds(m_{c}^{2}-s)^{2}}{s}e^{-s/M^{2}}  \notag \\
&&+\Pi _{\mathrm{NP}}(M^{2}).  \label{eq:DecayCF}
\end{eqnarray}%
The first term in Eq.\ (\ref{eq:DecayCF}) given by $s$ integral is the
perturbative component of $\Pi ^{\mathrm{OPE}}(M^{2},s_{0})$. The
nonperturbative term $\Pi _{\mathrm{NP}}(M^{2})$ has the following form%
\begin{eqnarray}
&&\Pi _{\mathrm{NP}}(M^{2})=-\frac{\langle \overline{d}d\rangle f_{K}m_{c}}{%
18}e^{-m_{c}^{2}/M^{2}}  \notag \\
&&+\langle \frac{\alpha _{s}G^{2}}{\pi }\rangle \frac{f_{K}m_{c}^{4}}{%
432M^{4}}\int_{0}^{1}\frac{dz}{z^{3}(z-1)^{3}}e^{-m_{c}^{2}/[M^{2}z(1-z)]}
\notag \\
&&+\frac{\langle \overline{d}g\sigma Gd\rangle f_{K}m_{c}^{3}}{72M^{4}}%
e^{-m_{c}^{2}/M^{2}}-\langle \frac{\alpha _{s}G^{2}}{\pi }\rangle \langle
\overline{d}d\rangle  \notag \\
&&\times \frac{f_{K}m_{c}(m_{c}^{2}+3M^{2})\pi ^{2}}{324M^{6}}%
e^{-m_{c}^{2}/M^{2}}+\langle \frac{\alpha _{s}G^{2}}{\pi }\rangle \langle
\overline{d}g\sigma Gd\rangle  \notag \\
&&\times \frac{f_{K}m_{c}(m_{c}^{4}+7M^{2}m_{c}^{2}+8M^{4})\pi ^{2}}{%
1296M^{10}}e^{-m_{c}^{2}/M^{2}}.  \label{eq:DecayNPCF}
\end{eqnarray}

\begin{table}[tbp]
\begin{tabular}{|c|c|}
\hline\hline
Parameters & Values (in $\mathrm{MeV}$ units) \\ \hline
$m_{D}$ & $1869.65\pm 0.05$ \\
$m_{D^0}$ & $1864.83\pm 0.05$ \\
$m_{K}$ & $493.677\pm 0.016$ \\
$m_{K^0}$ & $497.611\pm 0.013$ \\
$f_{D}=f_{D^0}$ & $212.6 \pm 0.7$ \\
$f_{K}=f_{K^0}$ & $155.7 \pm 0.3$ \\ \hline\hline
\end{tabular}%
\caption{Parameters of the $D$ and $K$ mesons used in numerical analyses.
Masses and decay constants of the mesons $\overline{D}^{0}$ and $K^{0}$ are
denoted by $m_{D^0}$, $f_{D^0}$ and $m_{K^0}$, $f_{K^0}$, respectively. }
\label{tab:Param}
\end{table}

The second decay $X_{1}\rightarrow \overline{D}^{0}K^{0}$ can be analyzed by
a same manner. Difference appears in the expression of the correlation
function (\ref{eq:CorrF6}), in which one should change the propagator $%
\widetilde{S}_{d}$ to $\widetilde{S}_{u}$, and the quark field $\overline{u}%
\rightarrow \overline{d}$. Related replacements $\langle \overline{d}%
d\rangle \rightarrow \langle \overline{u}u\rangle $, $\langle \overline{d}%
g\sigma Gd\rangle \rightarrow \langle \overline{u}g\sigma Gu\rangle $ in
Eq.\ (\ref{eq:DecayNPCF}) do not change numerical predictions.

Besides the vacuum condensates, Eq.\ (\ref{eq:SRcoupl}) contains masses and
decay constants of the final-state mesons $D^{-}$ and $K^{+}$. Their
spectroscopic parameters, as well as parameters of the mesons $\overline{D}%
^{0}$ and $K^{0}$ are collected in Table\ \ref{tab:Param}: All of them are
borrowed from Ref.\ \cite{PDG:2020}.

Our analysis demonstrates that working windows (\ref{eq:Wind1}) used in the
mass calculations satisfy necessary constraints on $M^{2}$ and $s_{0}$
imposed in the case of the decay process. Therefore, in computations of $\Pi
^{\mathrm{OPE}}(M^{2},s_{0})$ we vary $M^{2}$ and $s_{0}$ within limits (\ref%
{eq:Wind1}).

For $g_{1}$ numerical calculations yield
\begin{equation}
g_{1}=8.6\pm 1.1.  \label{eq:Coupl1}
\end{equation}%
The partial width of the decay $X_{1}\rightarrow D^{-}K^{+\text{ }}$ is
determined by the simple formula
\begin{equation}
\Gamma _{\mathrm{A}}\left[ X_{1}\rightarrow D^{-}K^{+}\right] =\frac{%
g_{1}^{2}\lambda ^{3}\left( m,m_{D},m_{K}\right) }{24\pi m^{2}},
\label{eq:DW}
\end{equation}%
where
\begin{eqnarray}
\lambda \left( a,b,c\right) &=&\frac{1}{2a}\left[ a^{4}+b^{4}+c^{4}\right.
\notag \\
&&\left. -2\left( a^{2}b^{2}+a^{2}c^{2}+b^{2}c^{2}\right) \right] ^{1/2}.
\end{eqnarray}%
Then it is not difficult to find that
\begin{equation}
\Gamma _{\mathrm{A}}\left[ X_{1}\rightarrow D^{-}K^{+}\right] =(46\pm 9)~%
\mathrm{MeV}.  \label{eq:DW1Numeric}
\end{equation}

The strong coupling and partial width of the second process $%
X_{1}\rightarrow \overline{D}^{0}K^{0}$ can be obtained by the same way:
\begin{eqnarray}
&&g_{2}=8.7\pm 1.1,  \notag \\
&&\Gamma _{\mathrm{B}}\left[ X_{1}\rightarrow \overline{D}^{0}K^{0}\right]
=(47\pm 9)~\mathrm{MeV}.  \label{eq:DWA}
\end{eqnarray}%
Differences between couplings $g_{1}$ and $g_{2}$, and partial widths of two
channels originate from parameters of the final-state mesons, therefore are
very small. Extracted predictions for $\Gamma _{\mathrm{A}}$ and $\Gamma _{%
\mathrm{B}}$ allow us to evaluate the full width of the tetraquark $X_{1}$%
\begin{equation}
\Gamma _{\mathrm{full}}=(93\pm 13)~\mathrm{MeV}.
\end{equation}%
Confronting the sum rule's prediction for $\Gamma _{\mathrm{full}}$ with the
LHCb data, one sees that $\Gamma _{\mathrm{full}}$ is smaller than the measured
value in Eq.\ (\ref{eq:Data2}). Nevertheless, within ambiguities of
theoretical calculations, $\Gamma _{\mathrm{full}}$ is in a reasonable
agreement with $\Gamma _{1}$.

%%%%%%%%%%%%%%%%%%%%%%%%%%%%%%%%%%%%%%%%%%%%%%%%%%%%%%%%%%%%%%%%%%%%%%%%%%%%

\section{Discussion and concluding notes}

\label{sec:Disc}
%%%%%%%%%%%%%%%%%%%%%%%%%%%%%%%%%%%%%%%%%%%%%%%%%%%%%%%%%%%
In the present paper, we have investigated the new resonance $X_{1}$
observed recently by the LHCb collaboration. We have modeled $X_{1}$ as the
vector diquark-antidiquark state $[ud][\overline{c}\overline{s}]$ and computed its
mass and full decay width $\Gamma _{\mathrm{full}}$. The mass of the state $%
X_{1}=[ud][\overline{c}\overline{s}]$ has been evaluated in the framework of
the QCD two-point sum rule approach. Obtained prediction for $m$ nicely
agrees with the LHCb data, and may be considered as arguments in favor of
diquark-antidiquark structure of the resonance $X_{1}$.

We have evaluated the full width of the tetraquark $X_{1}$ as well. To this
end, we have analyzed its two $P$-wave decay modes $X_{1}\rightarrow
D^{-}K^{+}$ and $X_{1}\rightarrow \overline{D}^{0}K^{0}$. Strong couplings
corresponding to vertices $X_{1}D^{-}K^{+}$ and $X_{1}\overline{D}^{0}K^{0}$
have been calculated by employing the LCSR method and soft-meson
approximation. It has been found that their widths do not differ from each
another, and these channels form the full width of the tetraquark $X_{1}$ on
equal footing. The result obtained for $\Gamma _{\mathrm{full}}$ is
compatible with LHCb measurements.

The mass of $X_{1}$ was calculated in the context of the sum rule method
also in Ref.\ \cite{Chen:2020aos}. The prediction $m=2940_{-110}^{+130}~\mathrm{MeV}$
obtained there is somewhat larger than our result, but within theoretical
errors still agrees with the LHCb data. Relatively large output for $m$ is
presumably connected with a form of interpolating current and accuracy of
performed calculations.

Production of the structures $X_{0}$ and $X_{1}$ in $B$ meson's weak decays
were analyzed in Ref.\ \cite{Burns:2020xne}. The central idea and main
conclusion of this work is that production of $X_{0,1}$ is dominated by
color-favored processes. It was also argued that competing models for $%
X_{0,1}$ can be unambiguously discriminated due to differences in features
of their production and decay mechanisms. Similar problems were addressed in
articles  \cite{Bondar:2020eoa,Chen:2020eyu} as well.

The resonances $X_{0,1}$ are neutral structures, but may have charged
partners \cite{Burns:2020xne}. In our view, more interesting is a case of
exotic mesons built of \ four quarks of different flavors and carrying two
units of the electric charge. We are going now to consider production of
doubly charged tetraquarks $Z^{++}=[uc][\overline{s}\overline{d}]$ in $B$
decays. In Ref. \cite{Agaev:2017oay} we investigated scalar, pseudoscalar
and axial-vector states $Z_{\overline{c}s}=[sd][\overline{u}\overline{c}]$
with the charge $-2|e|$, and computed their masses and decay widths.
Tetraquarks $Z^{++}$ \ are positively charged partners of $Z_{\overline{c}s}$
and should have the same masses. Therefore, in our analysis of $Z^{++}$, we
use results presented in Ref.\ \cite{Agaev:2017oay}. Then scalar and
axial-vector states $Z_{\mathrm{S}}^{++}$ and $Z_{\mathrm{AV}}^{++}$ should
have the masses $2628_{-153}^{+166}~\mathrm{MeV}$ and $2826_{-157}^{+134}~%
\mathrm{MeV}$, respectively. We did not calculate parameters of vector
tetraquark $Z_{\overline{c}s}$, but can safely suppose that $Z_{\mathrm{V}%
}^{++}$ is heavier than $Z_{\mathrm{S}}^{++}$ and its mass is comparable
with mass of $Z_{\mathrm{AV}}^{++}$. The scalar particle $Z_{\mathrm{S}%
}^{++} $ in $S$-wave can decay to mesons $D_{s}^{+}\pi ^{+}$ and $D^{+}K^{+}$%
. The vector tetraquark $Z_{\mathrm{V}}^{++}$ in $P$-wave has the same decay
modes. In other words, decays to ordinary meson pairs $D_{s}^{+}\pi ^{+}$
and $D^{+}K^{+}$ are kinematically allowed processes for both $Z_{\mathrm{S}%
}^{++}$ and $Z_{\mathrm{V}}^{++}$.

The structures $X_{0}$ and $X_{1}$ were discovered in the process $%
B^{+}\rightarrow D^{+}X\rightarrow D^{+}D^{-}K^{+}$ and fixed in the $%
D^{-}K^{+}$ invariant mass distribution. This decay runs through
color-favored and color-suppressed topologies labeled in Ref. \cite%
{Burns:2020xne} as $(1)$ and $(2)$, respectively. It is not difficult to see
that weak decays of $B^{+}$ with the same topologies may generate the
process $B^{+}\rightarrow D^{-}Z^{++}\rightarrow D^{-}D^{+}K^{+}$ as well.
As a result, doubly charged scalar and vector tetraquarks may manifest
themselves in the invariant mass distribution of the pair $D^{+}K^{+}$.
There is intriguing possibility to observe doubly charged four-quark
structures in decay $B^{+}\rightarrow D^{-}D^{+}K^{+}$: It is quite
possible, that decays through $X$ and $Z^{++}$ are competing mechanisms in
this process. Other $B$ meson channels with $D_{s}^{+}\pi ^{+}$ pairs in
final-state, perhaps, are suitable for such studies as well. Experimental
data collected by the LHCb collaboration would hopefully be enough to
perform relevant investigations.

As is seen, there are different interpretations of new structures discovered
recently by the LHCb collaboration in decay $B^{+}\rightarrow
D^{+}D^{-}K^{+} $. Experimental data do not raise doubts about existence of
the resonance-like enhancements $X_{0}$ and $X_{1}$ in the $D^{-}K^{+}$ mass
distribution. Till now $X_{0}$ and $X_{1}$ were examined as the hadronic
molecules, diquark-antidiquark systems, and rescattering effects. In our
view, the same decay \ $B^{+}\rightarrow D^{+}D^{-}K^{+}$ may also be used
to see doubly charged resonances. Additional experimental and theoretical
studies are evidently required to clarify all these problems.


\begin{thebibliography}{99}
%\cite{LHCb:2020A}

\bibitem{LHCb:2020A} R.~Aaij \textit{et al.} [LHCb],
%``Model-independent study of structure in  $B^+\to D^+D^-K^+$ decays,''
Phys.\ Rev.\ Lett. \textbf{125}, 242001 (2020).
%[arXiv:2009.00025 [hep-ex]].

%\cite{LHCb:2020}

\bibitem{LHCb:2020} R.~Aaij \textit{et al.} [LHCb],
%``Amplitude analysis of the $B^+\to D^+D^-K^+$ decay,''
Phys.\ Rev.\ D \textbf{102}, 112003 (2020).
%doi:10.1103/PhysRevD.102.112003
%[arXiv:2009.00026 [hep-ex]].

%\cite{Karliner:2020vsi}

\bibitem{Karliner:2020vsi} M.~Karliner and J.~L.~Rosner,
%``First exotic hadron with open heavy flavor: $cs\bar u\bar d$ tetraquark,''
Phys.\ Rev.\ D \textbf{102}, 094016 (2020). %arXiv:2008.05993 [hep-ph].
%10 citations counted in INSPIRE as of 19 Aug 2020

%\cite{Wang:2020xyc}

\bibitem{Wang:2020xyc} Z.~G.~Wang,
%``Analysis of the $X_0(2900)$ as the scalar tetraquark state via the QCD sum rules,''
Int.\ J.\ Mod.\ Phys.\ A \textbf{35}, 2050187 (2020).
%arXiv:2008.07833 [hep-ph].
%0 citations counted in INSPIRE as of 19 Aug 2020

%\cite{Chen:2020aos}

\bibitem{Chen:2020aos} H.~X.~Chen, W.~Chen, R.~R.~Dong and N.~Su,
%``$X_0(2900)$ and $X_1(2900)$: hadronic molecules or compact tetraquarks,''
Chin.\ Phys.\ Lett. \textbf{37}, 101201 (2020). %arXiv:2008.07516 [hep-ph].
%3 citations counted in INSPIRE as of 19 Aug 2020

%\cite{Liu:2020nil}

\bibitem{Liu:2020nil} M.~Z.~Liu, J.~J.~Xie and L.~S.~Geng,
%``$X_0(2866)$ as a $D^*\bar{K}^*$ molecular state,''
Phys.\ Rev.\ D \textbf{102}, 091502 (2020). %arXiv:2008.07389 [hep-ph].
%4 citations counted in INSPIRE as of 19 Aug 2020

%\cite{Molina:2020hde}

\bibitem{Molina:2020hde} R.~Molina and E.~Oset,
%``Molecular picture for the $X_0(2866)$ as a $D^* \bar{K}^*$ $J^P=0^+$ state and related $1^+,2^+$ states,''
Phys.\ Lett.\ B \textbf{811}, 135870 (2020). %arXiv:2008.11171 [hep-ph].
%0 citations counted in INSPIRE as of 26 Aug 2020

%\cite{Hu:2020mxp}

\bibitem{Hu:2020mxp} M.~W.~Hu, X.~Y.~Lao, P.~Ling and Q.~Wang,
%``The molecular nature of the $X_0(2900)$,''
arXiv:2008.06894 [hep-ph].
%3 citations counted in INSPIRE as of 19 Aug 2020

%\cite{He:2020jna}

\bibitem{He:2020jna} X.~G.~He, W.~Wang and R.~Zhu,
%``Open-charm tetraquark $X_c$ and open-bottom tetraquark $X_b$,''
Eur.\ Phys.\ J.\ C \textbf{80}, 1026 (2020). %arXiv:2008.07145 [hep-ph].
%3 citations counted in INSPIRE as of 19 Aug 2020

%\cite{Liu:2020orv}

\bibitem{Liu:2020orv} X.~H.~Liu, M.~J.~Yan, H.~W.~Ke, G.~Li and J.~J.~Xie,
%``Triangle singularity as the origin of $X_0(2900)$ and $X_1(2900)$ observed in $B^+\to D^+ D^- K^+$,''
arXiv:2008.07190 [hep-ph].
%3 citations counted in INSPIRE as of 19 Aug 2020

%\cite{Lu:2020qmp}

\bibitem{Lu:2020qmp} Q.~F.~Lu, D.~Y.~Chen and Y.~B.~Dong,
%``Open charm and bottom tetraquarks in an extended relativized quark model,''
Phys.\ Rev.\ D \textbf{102}, 074021 (2020). %arXiv:2008.07340 [hep-ph].
%3 citations counted in INSPIRE as of 19 Aug 2020

%\cite{Zhang:2020oze}

\bibitem{Zhang:2020oze} J.~R.~Zhang,
%``An open charm tetraquark candidate: note on $X_{0}(2900)$,''
arXiv:2008.07295 [hep-ph].
%3 citations counted in INSPIRE as of 19 Aug 2020

%\cite{Huang:2020ptc}

\bibitem{Huang:2020ptc} Y.~Huang, J.~X.~Lu, J.~J.~Xie and L.~S.~Geng,
%``Strong decays of $\bar{D}^{*}K^{*}$ molecules and the newly observed $X_{0,1}$ states,''
Eur.\ Phys.\ J.\ C \textbf{80}, 973 (2020). %arXiv:2008.07959 [hep-ph].
%0 citations counted in INSPIRE as of 19 Aug 2020

%\cite{Xue:2020vtq}

\bibitem{Xue:2020vtq} Y.~Xue, X.~Jin, H.~Huang and J.~Ping,
%``Tetraquarks with open charm flavor,''
arXiv:2008.09516 [hep-ph].
%0 citations counted in INSPIRE as of 24 Aug 2020

%\cite{Yang:2021izl}

\bibitem{Yang:2021izl} G.~Yang, J.~Ping and J.~Segovia,
%``The $\mathbf{sQ\bar{q}\bar{q}}$ $\mathbf{(q=u,\,d;\, Q=c,\,b)}$ tetraquarks in the chiral quark model,''
arXiv:2101.04933 [hep-ph].
%0 citations counted in INSPIRE as of 04 Mar 2021

%\cite{Wu:2020job}

\bibitem{Wu:2020job} T.~W.~Wu, M.~Z.~Liu and L.~S.~Geng,
%``Excited $K$ meson, $K_c(4180)$, with hidden charm as a $D\bar{D}K$ bound state,''
Phys.\ Rev.\ D \textbf{103}, L031501 (2021).
%doi:10.1103/PhysRevD.103.L031501
%arXiv:2012.01134 [hep-ph]].
%2 citations counted in INSPIRE as of 04 Mar 2021

%\cite{Abreu:2020ony}

\bibitem{Abreu:2020ony} L.~M.~Abreu,
%``$X_J(2900)$ states in a hot hadronic medium,''
Phys.\ Rev.\ D \textbf{103}, 036013 (2021).
%doi:10.1103/PhysRevD.103.036013
%[arXiv:2010.14955 [hep-ph]].
%2 citations counted in INSPIRE as of 04 Mar 2021

%\cite{Wang:2020prk}

\bibitem{Wang:2020prk} G.~J.~Wang, L.~Meng, L.~Y.~Xiao, M.~Oka and
S.~L.~Zhu,
%``Mass spectrum and strong decays of tetraquark $\bar c\bar s qq$ states,''
Eur.\ Phys.\ J.\ C \textbf{81}, 188 (2021).
%doi:10.1140/epjc/s10052-021-08978-0
%[arXiv:2010.09395 [hep-ph]].
%2 citations counted in INSPIRE as of 04 Mar 2021

%\cite{Xiao:2020ltm}

\bibitem{Xiao:2020ltm} C.~J.~Xiao, D.~Y.~Chen, Y.~B.~Dong and G.~W.~Meng,
%``Study of the decays of $S-$wave $\bar D^\ast K^\ast$ hadronic molecules: The scalar $X_0(2900)$ and its spin partners $X_{J(J=1,2)}$,''
Phys.\ Rev.\ D \textbf{103}, 034004 (2021).
%doi:10.1103/PhysRevD.103.034004
%[arXiv:2009.14538 [hep-ph]].
%5 citations counted in INSPIRE as of 04 Mar 2021

%\cite{Dong:2020rgs}

\bibitem{Dong:2020rgs} X.~K.~Dong and B.~S.~Zou,
%``Prediction of possible $DK_1$ bound states,''
arXiv:2009.11619 [hep-ph].
%3 citations counted in INSPIRE as of 04 Mar 2021

%\cite{Burns:2020xne}

\bibitem{Burns:2020xne} T.~J.~Burns and E.~S.~Swanson,
%``Discriminating among interpretations for the $X(2900)$ states,''
Phys.\ Rev.\ D \textbf{103}, 014004 (2021).
%doi:10.1103/PhysRevD.103.014004
%[arXiv:2009.05352 [hep-ph]].
%8 citations counted in INSPIRE as of 04 Mar 2021

%\cite{Bondar:2020eoa}

\bibitem{Bondar:2020eoa} A.~E.~Bondar and A.~I.~Milstein,
%``Charge asymmetry in decays $B \to \bar{D}DK$,''
JHEP \textbf{12}, 015 (2020). %doi:10.1103/PhysRevD.103.014004
%[arXiv:2009.05352 [hep-ph]].
%8 citations counted in INSPIRE as of 04 Mar 2021

%\cite{Chen:2020eyu}

\bibitem{Chen:2020eyu} Y.~K.~Chen, J.~J.~Han, Q.~F.~L\"{u}, J.~P.~Wang and
F.~S.~Yu,
%``Branching fractions of $B^-\rightarrow D^-X_{0,1}(2900)$ and their implications,''
Eur.\ Phys.\ J.\ C \textbf{81}, 71 (2021).
%doi:10.1140/epjc/s10052-021-08857-8
%[arXiv:2009.01182 [hep-ph]].
%7 citations counted in INSPIRE as of 04 Mar 2021

%\cite{Albuquerque:2020ugi}

\bibitem{Albuquerque:2020ugi} R.~M.~Albuquerque, S.~Narison,
D.~Rabetiarivony and G.~Randriamanatrika,
%``$X_{0,1}$(2900) and $(D^-K^+)$ invariant mass from QCD Laplace sum rules at NLO,''
Nucl.\ Phys.\ A \textbf{1007}, 122113 (2021).
%doi:10.1016/j.nuclphysa.2020.122113
%[arXiv:2008.13463 [hep-ph]].
%10 citations counted in INSPIRE as of 04 Mar 2021


%\cite{Chen:2021erj}
\bibitem{Chen:2021erj}
H.~X.~Chen,
%``Hadronic molecules in B decays,''
arXiv:2103.08586 [hep-ph].

%\cite{Agaev:2020nrc}

\bibitem{Agaev:2020nrc} S.~S.~Agaev, K.~Azizi and H.~Sundu,
% New scalar resonance....
arXiv:2008.13027 [hep-ph]. %%CITATION = ARXIV:1603.01471;%%

%\cite{Agaev:2016lkl}

\bibitem{Agaev:2016lkl} S.~S.~Agaev, K.~Azizi and H.~Sundu, Phys.\ Rev.\ D\
\textbf{93}, 094006 (2016).
%Charmed partner of the exotic $X(5568)$ state and its properties,
%arXiv:1603.01471 [hep-ph].
%%CITATION = ARXIV:1603.01471;%%

%\cite{Chen:2016mqt}

\bibitem{Chen:2016mqt} W.~Chen, H.~X.~Chen, X.~Liu, T.~G.~Steele and
S.~L.~Zhu,
%``decoding the $X(5568)$ as a fully open-flavor $su\bar b\bar d$ tetraquark state,''
Phys.\ Rev.\ Lett.\ \textbf{117}, 022002 (2016).
%arXiv:1602.08916 [hep-ph].  %%CITATION = ARXIV:1602.08916;%%

%\cite{Agaev:2017oay}

\bibitem{Agaev:2017oay} S.~S.~Agaev, K.~Azizi and H.~Sundu,
%``Testing the doubly charged charm–strange tetraquarks,''
Eur.\ Phys.\ J.\ C \textbf{78}, 141 (2018).
%doi:10.1140/epjc/s10052-018-5640-4  [arXiv:1710.01971 [hep-ph]].  %%CITATION = doi:10.1140/epjc/s10052-018-5640-4;%%  %3 citations counted in INSPIRE as of 10 Mar 2018

%\cite{Shifman:1978bx}

\bibitem{Shifman:1978bx} M.~A.~Shifman, A.~I.~Vainshtein and V.~I.~Zakharov,
%``QCD and Resonance Physics. Theoretical Foundations,''
Nucl.\ Phys.\ B \textbf{147}, 385 (1979).
%doi:10.1016/0550-3213(79)90022-1  %%CITATION = doi:10.1016/0550-3213(79)90022-1;%%  %4985 citations counted in INSPIRE as of 11 Apr 2018

%\cite{Shifman:1978by}

\bibitem{Shifman:1978by} M.~A.~Shifman, A.~I.~Vainshtein and V.~I.~Zakharov,
%``QCD and Resonance Physics: Applications,''
Nucl.\ Phys.\ B \textbf{147}, 448 (1979).
%doi:10.1016/0550-3213(79)90023-3  %%CITATION = doi:10.1016/0550-3213(79)90023-3;%%  %2767 citations counted in INSPIRE as of 11 Apr 2018

%\cite{Chen:2016qju}

\bibitem{Chen:2016qju} H.~X.~Chen, W.~Chen, X.~Liu and S.~L.~Zhu,
%``The hidden-charm pentaquark and tetraquark states,''
Phys.\ Rept.\ \textbf{639}, 1 (2016). %doi:10.1016/j.physrep.2016.05.004
%[arXiv:1601.02092 [hep-ph]].

%\cite{Chen:2016spr}

\bibitem{Chen:2016spr} H.~X.~Chen, W.~Chen, X.~Liu, Y.~R.~Liu and S.~L.~Zhu,
%``A review of the open charm and open bottom systems,''
Rept.\ Prog.\ Phys.\ \textbf{80}, 076201 (2017).
%doi:10.1088/1361-6633/aa6420
%[arXiv:1609.08928 [hep-ph]].

%\cite{Albuquerque:2018jkn}

\bibitem{Albuquerque:2018jkn} R.~M.~Albuquerque, J.~M.~Dias,
K.~P.~Khemchandani, A.~Mart\'\i{}nez Torres, F.~S.~Navarra, M.~Nielsen and
C.~M.~Zanetti, %``QCD sum rules approach to the $X,~Y$ and $Z$ states,''
J.\ Phys.\ G \textbf{46}, 093002 (2019). %doi:10.1088/1361-6471/ab2678
%[arXiv:1812.08207 [hep-ph]].

%\cite{Agaev:2020zad}

\bibitem{Agaev:2020zad} S.~S.~Agaev, K.~Azizi and H.~Sundu,
%``Four-quark exotic mesons,''
Turk.\ J.\ Phys.\ \textbf{44}, 95 (2020). %%CITATION = ARXIV:2004.12079;%%

%\cite{Kondo:2004cr}

\bibitem{Kondo:2004cr} Y.~Kondo, O.~Morimatsu and T.~Nishikawa,
%``Two-hadron-irreducible QCD sum rule for pentaquark baryon,''
Phys.\ Lett.\ B \textbf{611}, 93 (2005).
%doi:10.1016/j.physletb.2005.01.070
%[arXiv:hep-ph/0404285 [hep-ph]].

%\cite{Lee:2004xk}

\bibitem{Lee:2004xk} S.~H.~Lee, H.~Kim and Y.~Kwon,
%``Parity of Theta+(1540) from QCD sum rules,''
Phys.\ Lett.\ B \textbf{609}, 252 (2005).
%doi:10.1016/j.physletb.2005.01.029
%[arXiv:hep-ph/0411104 [hep-ph]].

%\cite{Wang:2015nwa}

\bibitem{Wang:2015nwa} Z.~G.~Wang,
%``Analysis of the $Z_c(4200)$ as axial-vector molecule-like state,''
Int.\ J.\ Mod.\ Phys.\ A \textbf{30}, 1550168 (2015).
%doi:10.1142/S0217751X15501687
%[arXiv:1502.01459 [hep-ph]].

%\cite{Agaev:2018vag}

\bibitem{Agaev:2018vag} S.~S.~Agaev, K.~Azizi, B.~Barsbay and H.~Sundu,
%``The doubly charmed pseudoscalar tetraquarks $T_{cc;\bar{s} \bar{s}}^{++}$ and $T_{cc;\bar{d} \bar{s}}^{++}$,''
Nucl.\ Phys.\ B \textbf{939}, 130 (2019).
%doi:10.1016/j.nuclphysb.2018.12.021
%[arXiv:1806.04447 [hep-ph]].

%\cite{Sundu:2018nxt}

\bibitem{Sundu:2018nxt} H.~Sundu, S.~S.~Agaev and K.~Azizi,
%``New charged resonance $Z_{c}^{-}(4100)$: the spectroscopic parameters and width,''
Eur.\ Phys.\ J. C \textbf{79}, 215 (2019).
%doi:10.1140/epjc/s10052-019-6737-0
%[arXiv:1812.10094 [hep-ph]].

%\cite{Ioffe:2005ym}

\bibitem{Ioffe:2005ym} B.~L.~Ioffe, %``QCD at low energies,''
Prog.\ Part.\ Nucl.\ Phys. \textbf{56}, 232 (2006).
%doi:10.1016/j.ppnp.2005.05.001
%[arXiv:hep-ph/0502148 [hep-ph]].

%\cite{Narison:2015nxh}

\bibitem{Narison:2015nxh} S.~Narison,
%``Decay Constants of Heavy-Light Mesons from QCD,''
Nucl.\ Part.\ Phys.\ Proc.\ \textbf{270-272}, 143 (2016).
%doi:10.1016/j.nuclphysbps.2016.02.030 [arXiv:1511.05903 [hep-ph]].
%%CITATION = doi:10.1016/j.nuclphysbps.2016.02.030;%%
%2 citations counted in INSPIRE as of 15 May 2016

%\cite{Balitsky:1989ry}

\bibitem{Balitsky:1989ry} I.~I.~Balitsky, V.~M.~Braun and
A.~V.~Kolesnichenko,
%``Radiative Decay Sigma+ ---> p gamma in Quantum Chromodynamics,''
Nucl.\ Phys.\ B \textbf{312}, 509 (1989). %doi:10.1016/0550-3213(89)90570-1
%519 citations counted in INSPIRE as of 24 Aug 2020

%\cite{Belyaev:1994zk}

\bibitem{Belyaev:1994zk} V.~M.~Belyaev, V.~M.~Braun, A.~Khodjamirian and
R.~Ruckl, %``D* D pi and B* B pi couplings in QCD,''
Phys.\ Rev.\ D \textbf{51}, 6177 (1995). %doi:10.1103/PhysRevD.51.6177
%[arXiv:hep-ph/9410280 [hep-ph]].
%483 citations counted in INSPIRE as of 24 Aug 2020

%\cite{Ioffe:1983ju}

\bibitem{Ioffe:1983ju} B.~L.~Ioffe and A.~V.~Smilga,
%``Nucleon Magnetic Moments and Magnetic Properties of Vacuum in QCD,''
Nucl.\ Phys.\ B \textbf{232}, 109 (1984). %doi:10.1016/0550-3213(84)90364-X
%472 citations counted in INSPIRE as of 24 Aug 2020

%\cite{Agaev:2016ijz}

\bibitem{Agaev:2016ijz} S.~S.~Agaev, K.~Azizi and H.~Sundu,
%``Width of the exotic $X_b(5568)$ state through its strong decay to $B_s^{0} \pi^{+}$,''
Phys.\ Rev.\ D \textbf{93}, 114007 (2016).
%doi:10.1103/PhysRevD.93.114007  [arXiv:1603.00290 [hep-ph]].  %%CITATION = doi:10.1103/PhysRevD.93.114007;%%  %35 citations counted in INSPIRE as of 01 Sep 2017

%\cite{Agaev:2016dev}

\bibitem{Agaev:2016dev} S.~S.~Agaev, K.~Azizi and H.~Sundu,
%Strong $Z_c^{+}(3900)\rightarrow J/\psi \pi^{+}; \eta_{c} \rho^{+}$ decays in QCD,
Phys.\ Rev.\ D\ \textbf{93}, 074002 (2016). %\cite{Kondo:2004cr}

%\cite{PDG:2020}

\bibitem{PDG:2020} P.~A.~Zyla \textit{et al.} [Particle Data Group], Prog.\
Theor.\ Exp.\ Phys.\ \textbf{2020}, 083C01 (2020).
\end{thebibliography}
\end{document}